# Critical Assemblies: Dragon Burst Assembly & Solution Assemblies

Robert Kimpland, Travis Grove, Peter Jaegers, Richard Malenfant, and William Myers[*]
Los Alamos National Laboratory
Los Alamos, NM 87545

**Abstract:** This work reviewed the historical literature associated with the Dragon experiment and Water Boiler reactors operated at Los Alamos during the Manhattan Project. Frisch's invited talk given at the Fast Burst Reactor Conference held the University of New Mexico in Albuquerque, NM in 1969 is quoted. From the literature review, basic models for the Dragon experiment and for a Water Boiler type assembly (aqueous homogeneous reactor) were created that can be used for conducting multi-physics simulations for criticality excursion studies. This methodology utilizes the coupled neutronic-hydrodynamic method to perform a time-dependent dynamic simulation of a criticality excursion. MCNP® was utilized to calculate important nuclear kinetic parameters that were incorporated into the models. Simulation results compared reasonably well with historic data.

**Keywords:** Manhattan Project; Dragon Experiment; Water Boiler Reactors; Multiphysics Simulation; Coupled Neutronic-Hydrodynamic Method.

## Introduction

Lise Meitner and her nephew, Otto Frisch, discovered nuclear fission and named it such in a February 1939 issue of Nature. Physicists throughout the world immediately realized that the emission of 2-3 neutrons from the process could lead to a chain reaction and the release of a tremendous amount of energy.

The basic characteristics of a self-sustaining chain reaction were demonstrated with the Chicago Pile[1] in 1942, but it was not until early 1945 that sufficient enriched material became available to experimentally verify fast-neutron cross-sections and the kinetic characteristics of a nuclear chain reaction sustained with prompt neutrons alone.

Prompt neutrons are emitted by fission products within $10^{-12}$ seconds, and "delayed" neutrons are emitted a short time later. Reactors constructed in early 1945 including CP1 (the Chicago Pile)[1], the X-10 Reactor[2] at Oak Ridge, and the water boilers[3] at Los Alamos, all operated in a state sustained by delayed neutrons. If the neutron multiplication factor were increased so much that the prompt neutrons alone could support a divergent chain, the neutron population would grow exponentially, doubling in a small fraction of a second. A nuclear weapon would have to operate in this region.

During the Manhattan project at Los Alamos, the Dragon assembly and the Water Boiler assemblies played major roles as experimental capabilities to validate theory used to predict the dynamic behavior of neutron chain reacting systems. Both were used as a source of neutrons to acquire and validate nuclear data used by the theoretical models. The experimenters gained a wealth of data and experience on reactor operations that still form the basis of nuclear operations practiced today.

## History and Design Description of the Dragon Assembly

In October 1944, Otto R. Frisch presented a proposal to J. Robert Oppenheimer for a daring experiment to verify criticality based on prompt neutrons alone. Oppenheimer approved the proposal, and it was presented to the Coordinating Council. The Council included Enrico Fermi and Richard Feynman who likened the experiment to "tickling the tail of the sleeping dragon." Frisch's group began the design of the experiment and the machine named The Dragon[4] after Feynman's offhand comment. The experiment, elegantly simple in concept, consisted of dropping a slug of uranium hydride through a slightly sub-critical annulus of uranium hydride with a tamper of tungsten carbide or beryllium. For a small fraction of a second, a prompt neutron chain reaction was indicated by a burst of gamma-rays.

The Dragon was designed to support a chain reaction on prompt neutrons alone for about 1/100th second. By means of a series of static sub-critical experiments gradually approaching the critical state, the condition for the production of a reaction on prompt neutrons was defined. The static experiments also provided information on the reactivity worth of the slug and changes to the system's neutron multiplication as it traversed the annulus. With this information, the reactivity could be increased, and dynamic experiments performed.

[*] corresponding author: bmyers@lanl.gov

The first experiments were performed with a beryllium-oxide tamper. As more uranium hydride became available, it was decided to replace the beryllium with tungsten carbide to eliminate the neutron moderation resulting from beryllium. It was a surprise to find that the n-2n reaction in beryllium made a substantial contribution to the neutron economy.

It became clear that the magnitude of the burst of neutrons was determined by the initial neutron population. As a result, the slug of uranium hydride was returned to the top of the tower and dropped repeatedly to produce larger and larger bursts. The build up of fission products from successive drops, a fraction of which decay and emit neutrons were responsible for the increase of the intrinsic neutron source population, enabled larger bursts. In this way, bursts over $10^{15}$ neutrons were produced within 3 milliseconds resulting in heating rates of 2,000 degrees Celsius per second that corresponds to a peak power of 20,000 KW. A side benefit of this work was that the Dragon assembly was used as a tool to study delayed neutrons and delayed gamma rays. It provided the first opportunity to acquire nuclear data associated with fission product production caused by irradiation from a short intense burst[5,6].

Here are the words of Otto Frisch describing the Dragon experiment as transcribed from a keynote speech given at the American Nuclear Society's Fast Burst Reactor Conference held the University of New Mexico in Albuquerque, NM in 1969. Figure 1 is Otto Frisch's Manhattan project badge photo.

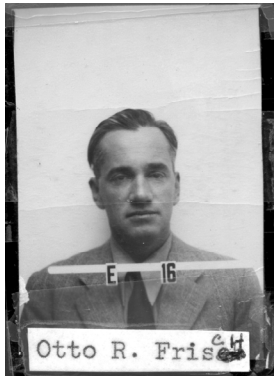

Figure 1. Otto Frisch's Manhattan project Badge photo at Los Alamos.

### *Frisch's Account of the Dragon Experiment[7]*

Just about 8 months ago my wife brought me a letter. I had overslept and was still rather sleepy, and I stared at the letter from the American Nuclear Society, National Topical Meeting, Fast Burst Reactors. And I said, "Do the Americans run a topical meeting on a nationwide scale because their reactors burst so fast?" After that I realized that I was being honored as the father, or I feel more like the grandfather, of pulsed reactors, having arranged the experiment which for the first time established a short-lasting harmless fast fission reaction that did not depend on delayed neutrons.

A group of no less than 17 people worked on this first controlled fission experiment known as the Dragon experiment, and I want to tell you how it came about.

The purpose of Los Alamos was to assemble part of the scientists who were needed to develop an atomic bomb. In particular, we measured the cross sections, time constants, and so on, which would make it possible to design a bomb with a reasonable degree of efficiency and safety. One of the most difficult things was to determine that the fast reaction would really work as fast as the theory predicted. Nuclear theory, of course, said that once a neutron hits a uranium nucleus, fission follows almost instantaneously, if it follows at all. But electronic methods at that time were not really fast enough to decide whether it happens with the sort of subnanosecond speed which was theoretically foreseen and needed if the bomb was to be an effective explosive.

So a number of ingenious experiments were devised to test the speed of the fission reaction, and the limit was pushed fairly well toward the point where we wanted it. But even so, I for one thought it would be very nice to go one step nearer to a real atomic explosion. It is a bit like the curiosity of the explorer who has climbed a volcano and wants to take one step nearer to look down into the crater but not fall in! That chance came when we learned that around the beginning of 1945 some amounts of separated $^{235}$U were to arrive. These shipments were meant mainly for us to carry out critical experiments to check the calculations of the theoreticians. The theoreticians, of course, had taken all the cross-section measurements—fission cross sections, elastic, inelastic, everything that we could produce for them—and from these, by complicated integrations, had worked out the critical size. However, experimental confirmation was desirable.

It was clear that we would not be able to test with a critical assembly of metallic $^{235}$U or metallic plutonium because once such a quantity had been produced the military would want to use it immediately. Instead, the first amount of $^{235}$U that came out of the mass separators was made into hydride, $UH_3$, and combined with a plastic binder into bricks of the approximate composition $UH_{10}$. The $^{235}$U enrichment was in the range of 71 to 75 weight percent[4]. You may ask why use that material; it would never make a useful bomb. That is quite true; but it enabled us to carry out critical measurements and compare them with calculations the theoreticians had performed for the same material. This comparison gave the theoreticians at least an idea about how reliable their calculations were and by how much and in which direction they might have to be corrected.

A large number of critical measurements were made, indeed, and the theoreticians were very pleased to have this corroboration of their calculations. In addition, I felt that here was a chance of looking a bit closer at the occurrence of a fast reaction, a reaction not limited by thermal neutrons, and I made the proposal that we should make an assembly with a hole in the middle, and that the missing portion should then be allowed to drop through the assembly under such conditions that for a few milliseconds the whole assembly would be critical with respect to prompt neutrons. I did a few simple calculations to be sure that this would be feasible, then sent this proposal to the coordinating council. Of course, I was not present when the proposal was discussed, but it was accepted; it was said that Enrico Fermi nodded his head in a pleased manner and said this was a nice experiment that we ought to try, and I was told that Dick Feynman, who was present, started to chuckle and to say that this is just like tickling the tail of a sleeping dragon. That is how the experiment was named.

When the $^{235}$U arrived, we built the equipment for the experiment, and Figure 2 roughly shows what this equipment looked like. It looks, crudely speaking, like an oil derrick, but it was only something like 6 m high. Near the bottom the uranium assembly was set upon a steel table. The material was available in the form of little bricks; I believe they were 1 in. by ½ in. by ½ in and very accurately made. (it was a joy to build little skyscrapers out of uranium hydride and other materials like that!) A slightly askew box that contained part of the assembly was mounted on a hydraulic pusher rod so it could be released and lowered—deliberately, rather slowly. The guides for the falling slug can also be seen.

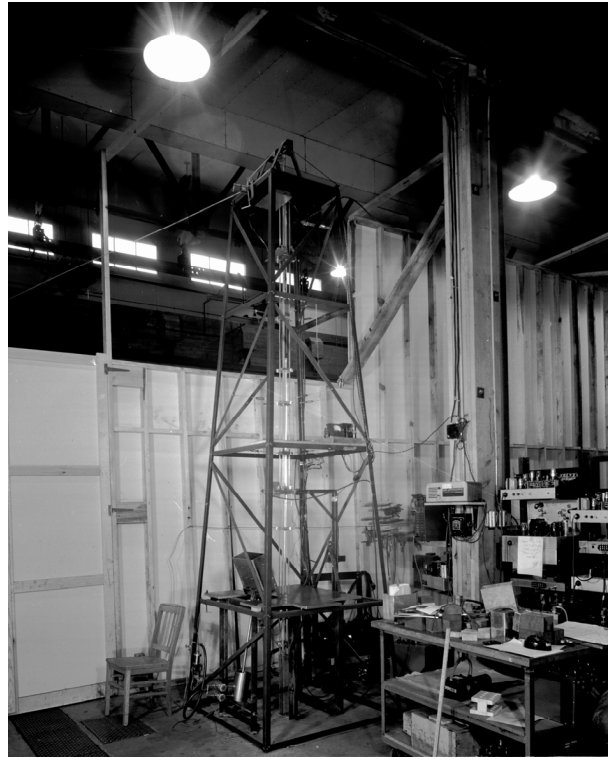

Figure 2. The Dragon Assembly Support Structure

Everybody, of course, asked me what if the slug gets stuck—you all blown up. We could not all be blown up. The material became only very slightly supercritical. It would have been a bomb of extremely low efficiency, and probably we would have been wise to clear out fast if the slug had stuck.

The top of the derrick contained a fairly elaborate device for holding the slug until everything was ready for the drop because we were aware that a danger much greater than the slug's getting stuck was the danger of the slug's falling before the supercriticality had been correctly adjusted (see Figure 3). We made quite sure that the slug could only be dropped after the operator had checked a certain number of things and was convinced that they were okay. In the end, of course, a great responsibility did fall on the operator.

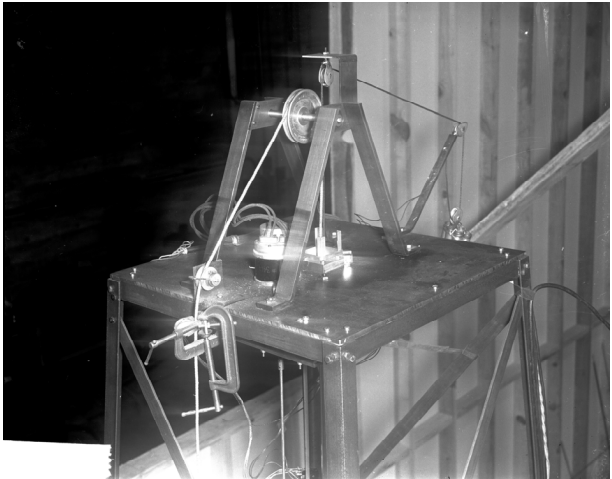

Figure 3. Top of the Dragon Assembly Support Structure

The steel table (about a centimeter thick) was placed so that material could rest firmly against the fuel box, which in actual use would be pushed up until it would be leaning against the guides or almost touching them. The gadgets attached to the guides measured the speed of the slug. You may say there is no reason to measure the speed if by some chance extra friction stops or slows the slug. The purpose of the measuring, however was to do a few dummy drops before beginning the day's work to make sure the slug was dropping according to Galileo's law. In fact it never did. It was always about 1% slower owing to friction. It did not fall freely in the guide; in fact, we deliberately leaned the whole tower a little to one side so that the slug was sliding down the guides rather than falling through them. This development was important because a very small sideways movement changed the multiplication constant and made a very big difference in the size of burst produced.

Much of the top of the assembly was crude and primitive. Parts were held together by ordinary mechanics clamps and there was a rope going up over several pulleys and holding an electromagnet that hauled up the slug. The electromagnet could not be switched off until after everything else was straightened out. I will not bore you with the safety precautions; they are completely out of date. What I really wanted to impress upon you is the rather primitive setup. This entire reactor was built in a matter of a few weeks, and all the experiments were performed during, I believe, three short periods in three weeks, each lasting only a few days. The reason we worked so fast was that the chemists were waiting for us to return the material so that together with further $^{235}$U it could be turned into metal and this material into bombs as soon as possible.

With the very first material that arrived, we made a number of drops to make sure the device worked, and the first pulses were obtained just at this time of year 24 years ago. Then we replaced the material with a somewhat bigger assembly and performed a number of drops to test the theory, and they all came out as we expected. The pulses were of the duration that we had approximately predicted from nuclear data. Figure 4 shows the outcome obtained by having a boron chamber close to the arrangement, which was connected to a cathode-ray oscilloscope, and simply integrated the amount of charge deposited. The figure reads from left to right in units of 6 msec, the rate at which the oscilloscope was pulsed. The charge suddenly begins to increase, increase more rapidly, and then straighten out once again; this is the integrated pulse. This result could be compared with the theory, and the agreement was very good.

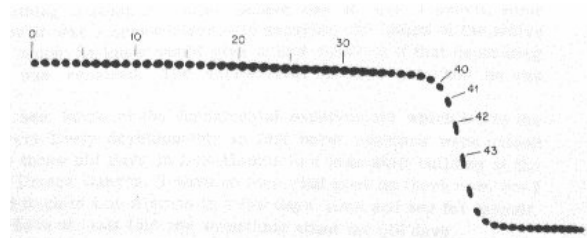

Figure 4. Integrated pulse reading where the graph's abscissa represents time plotted left to right in units of 6 msec.

Well, so much for the Dragon. It was dismantled a few weeks after it had been built, and the whole group dispersed.

We did perform a few Dragon-type experiments of the more modern variety unintentionally and with tragic results. Two men who worked on the Dragon experiment were killed within a few months after this experiment. Harry Daghlian, working by himself one night, which was breaking the rules, did not know how slippery the large blocks of tungsten carbide reflecting material were. While trying to put on one more block, he realized that the reaction was going up much too fast; he tried to pull the block away, but it slipped out of his hands. What then happened can only be reconstructed by theory; no one else was present. He saw a blue flash, and about 10 days later he died in the hospital from radiation damage. He had received well over a fatal dose. Probably what happened is that the material expanded thermally and thereby switched itself off, but the amount of radiation it had given off in that short time was enough.

Later I left Los Alamos and Louis Slotin took over the group working on critical assemblies. He told me that Fermi warned him, "You know that in this sort of work you have perhaps an even chance to survive your work here." Slotin was rather shaken about it. Even so, he did use something makeshift—some people say it was a pencil, some people say it was a screwdriver—to separate two lumps of the active material which he knew would give a fast reaction if that separating material was removed. The screwdriver slipped out, and he was killed.

So you see, some of the fundamental experiments which led to the present very lively developments in fast burst reactors were indeed started in those old days in Los Alamos in a makeshift building at the bottom of Omega Canyon. I have no idea what goes on there now, but I am hoping to go to Los Alamos in a few days' time and see for myself. Anyhow, I have at least told you something about the old days.

**History and Design Descriptions of the Water Boiler Reactors at Los Alamos**

For more than thirty years aqueous homogenous (nuclear) reactor (AHR) systems were designed, built, and operated at Los Alamos, starting in 1943 with the LOPO reactor and ending in 1974 with the SUPO reactor. These systems, also known as "water boilers", were typically enriched uranium solutions surrounded by a thick reflector and were utilized for critical mass studies, allowed measurement of neutron cross sections of various materials, provided valuable experience operating chain-reacting systems, and provided a source of fission neutrons for experimental studies.

The name of "water boiler" originated because at high power the systems seemed to "boil" due to bubbles of hydrogen and oxygen gas escaping from the system. This gas was generated through the radiolysis of the water in the solution, as the by-products of the fission process (radiation and fission products) would dissociate the water molecules into hydrogen and oxygen gases.

In 1944 the very first water boiler at Los Alamos was built and designed and given the name LOPO[8] ("low power"). See Figure 5. LOPO was the world's third artificial nuclear reactor (the first being Chicago Pile-1 at the University of Chicago in Illinois[9] and the second being the X-10 Pile at Oak Ridge, Tennessee[10]) and was the first reactor to be fueled by enriched uranium as well as the first reactor to use a liquid fuel. LOPO utilized a uranyl sulfate solution as fuel (14% enriched in uranium-235) and was surrounded by a thick beryllium oxide reflector as well as a graphite reflector. Uranyl sulfate was chosen instead of uranyl nitrate due to the lower neutron absorption in the sulfate as well as concerns about the solubility of the nitrate[11,12]. LOPO essentially operated at zero power, and after experimenters had performed the critical mass measurements and gotten valuable experience operating a solution reactor, a new reactor was designed and built using the same concept as LOPO but able to operate at higher power levels.

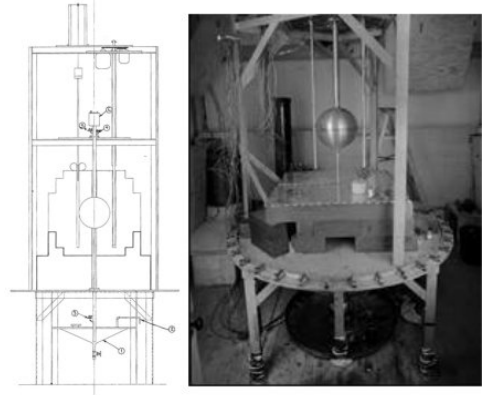

Figure 5. Schematic and photograph of LOPO.

This new reactor was named HYPO[13] ("high power") and was in operation in Los Alamos by late 1944. See Figures 6 and 7. HYPO was designed not only to operate at higher power levels (approximately 1-5 kilowatts) but to also provide a platform for more intricate and flexible experimental studies.

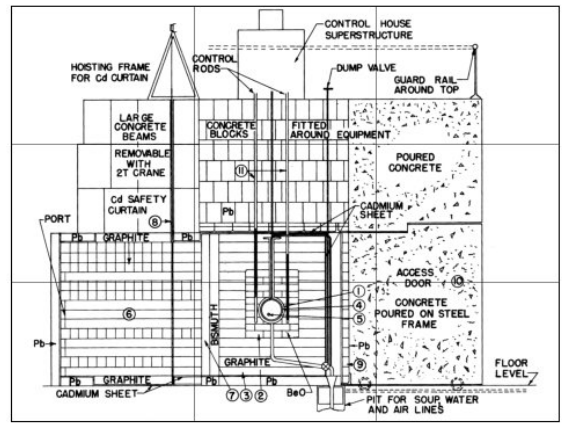

Figure 6. Facility Schematic of HYPO.

Additions to the Lopo design included adding cooling coils to provide a way to cool the solution fuel, additional control rods, a "glory hole" pipe that ran through the center of the core and provided experimental access, as well as a gamma-ray and neutron shield for personnel protection. Instead of using uranyl sulfate for fuel, HYPO used uranyl nitrate (14 weight % enriched in uranium-235) and was still reflected by a thick beryllium oxide reflector and a graphite reflector[11].

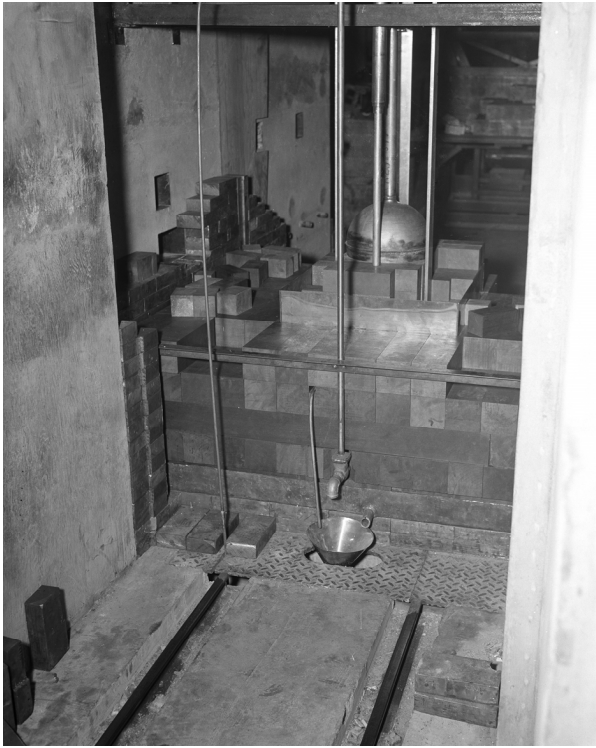

Figure 7. Photograph of HYPO.

In the late 1940's, higher neutron fluxes were deemed necessary for the experiments of the time, and in 1949 modifications began on HYPO to increase the ability to operate at higher powers (and thus higher neutron fluxes). These modifications included increasing the uranium enrichment in the uranyl nitrate fuel from 14 weight % to 89 weight %, the installation of larger cooling coils, the addition of a gas recombination system (so that the hydrogen and oxygen gas released in the radiolysis of the water could be recombined and could avoid any explosive hazards from these two gases), and finally the beryllium oxide portion of the reflector was replaced with graphite[11,14].

This new water boiler, named SUPO[15] ("super power"), began operation in 1950 (see Figures 8 and 9), and allowed operations up to 35 kilowatts, but were limited to steady state operations (normal mode) of 25 kw. At higher powers, the catalytic converter (part of the gas recombination system) would heat up and fail to keep up with radiolytic gas production causing the presence of detonable quantities of hydrogen and oxygen. Supo operated almost daily until it was finally shutdown for good in 1974. Over almost twenty-five years of near-daily operations, SUPO provided immense support for the Los Alamos weapons program (cross section measurements, fission yield calculations, and others), provided support for numerous health physics applications that examined the health effects of gamma-ray and neutron radiation, as well as transient behavior studies of solution reactor systems[11,14].

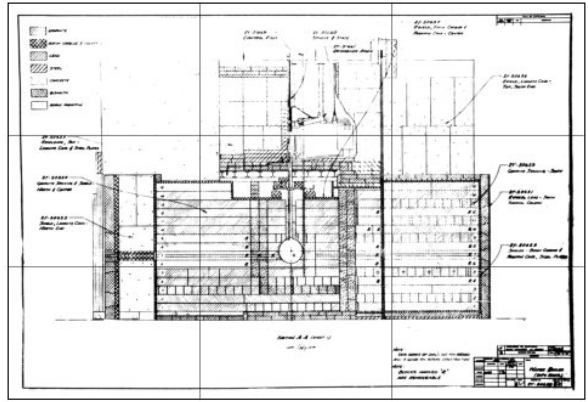

Figure 8. Facility Schematic of SUPO.

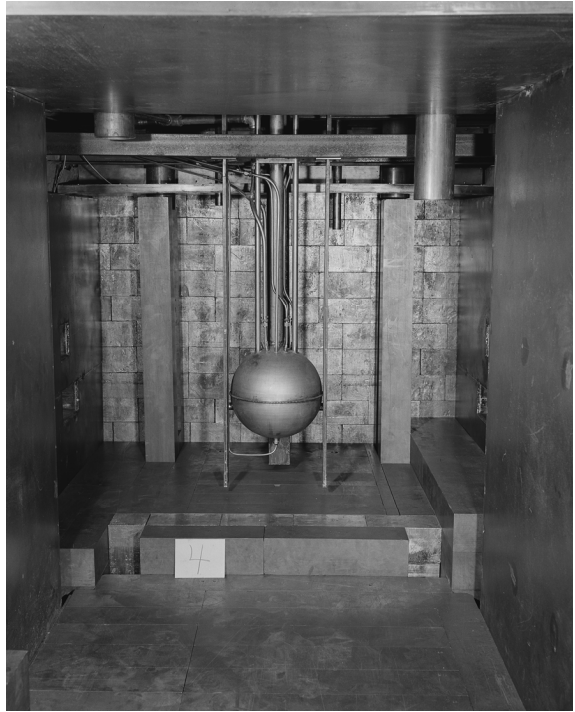

Figure 9. Photograph of SUPO Core.

## History of Criticality Excursion Studies at Los Alamos Using the Coupled Neutronic-Hydrodynamic Method

Los Alamos National Laboratory has a rich history of developing multi-physics simulation tools to study criticality excursions. The design philosophy for these simulation tools were based upon the "coupled neutronic-hydrodynamic method[16]". Development of the technique proceeded only as rapidly as reliable computers became available and it was not until the early 1950's under the guidance of Ernest W. Salimi and Conrad Longmire of the Los Alamos Scientific Laboratory that working codes were created to couple together the differential equations for thermodynamics, material motion, and neutron transport with rapidly changing reactivity.

The first non-weapon application of this technique was in 1957 subsequent to the accidently large power excursion in the original Lady Godiva[17] reactor using the code referred to as the "detailed method". This code and a later version known as the RAC code[18] were replaced in the early 1970's by the Los Alamos Pajarito Dynamics code known by the acronym PAD[19].

The PAD code employed the coupled neutronic-hydrodynamics method in one-dimension, with the neutronics provided by the DTF-IV[20] transport calculations. Due to difficulties of running the PAD code on more modern computers, the MRKJ Reactor Transient code[21] was developed.

The first version of the MRKJ code was developed for one-dimensioal geometries using the discrete ordinates neutron transport code ONEDANT[22] to perform the neutron transport calculations. The inferred data from the Lady Godiva accident excursion was used to help verify and validate the code[23]. This version of the code was used during a study of a criticality accident scenario associated with the proposed Yucca Mountain spent nuclear fuel storage facility[24].

A later version of the MRKJ code was developed for two-dimensional geometries[25] and used the PARTISN code[26] to perform the neutron transport calculations. Recent multi-physics modeling efforts[27,28] have revisted the Lady Godiva reactor accident excursions. This work utilized the Monte Carlo Application ToolKit (MCATK)[29,30,31] to perform the neutron transport calculations for the simulations.

**Dragon Criticality Excursion Simulations and Results**

To begin the multiphysics modeling process, one needs to create a model based upon the available experimental data. Figure 10 is a sketch[5] of the Dragon core that was used as a basis for creating a model to begin the simulation.

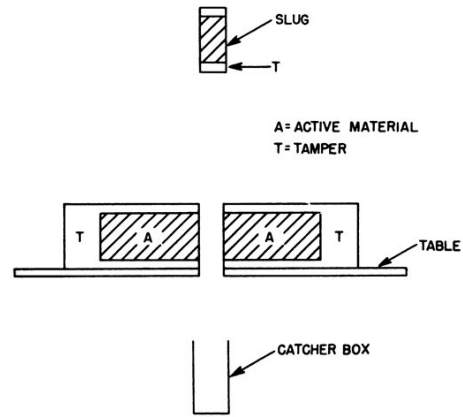

Figure 10: A Sketch of the Dragon assembly core.

*Dragon Neutronic Model*

Given the available data, a neutronic model of Dragon Assembly 2 was developed and analyzed using MCNP6.2[32]. Figure 11 illustrates a MCNP model representation of the Dragon core that was created to perform the neutronics calculations. The goal of this rudimentary analysis was to determine the possible range of operating parameters for the assembly. These parameters include the dimensions of the core, the multiplication factor of the assembly prior to a transient experiment (sans slug), and the potential reactivity worth of the slug. Assembly 2 was reported to have a core composed of 15.4 kg of $UH_{10}$ fuel. The fuel slug traveled in a steel box 2.125" x 2.125" x 14", thus the core had a rectangle cavity with at least the dimensions of the steel box. Assembly 1 reportedly had a 6" thick BeO tamper, so it was assumed that Assembly 2 had the same. It was also reported that Assembly 2 had a Cd layer surrounding the core, so it was assumed that this layer was 1/16" thick. Finally, the $UH_{10}$ fuel, which was a composite material of $UH_3$ and polystyrene, was assumed to have a density of 3.9 g/cm$^3$. With these considerations in mind, a proposed neutronic model consisting of a $UH_{10}$ core in the shape of a cube and surrounded by 6" of BeO along all vertical faces was analyzed. The core is 16.5 cm on a side with a 2.25" x 2.25" cavity in the center. A 0.5 cm thick layer of BeO was also placed along the horizontal surfaces of the core to account for reflection from various structural components. The base case slug was assumed to be a solid rectangular piece 2" x 2" with a length of 16.5 cm. This model yields a multiplication factor of 0.94277 sans slug and 1.02739 with the slug present in its maximum reactivity worth position. A generation time of approximately 1.1 microseconds was predicted for this proposed neutronic model, suggesting that it possesses roughly the same neutronic characteristics of the experiment.

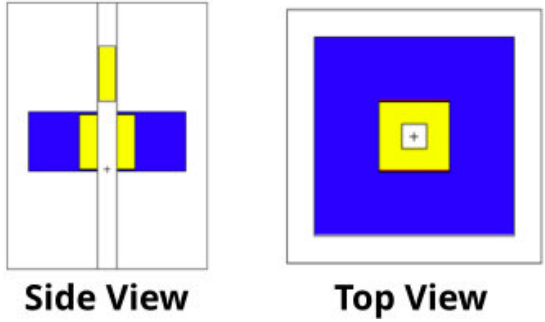

Figure 11. MCNP Representation of the Dragon Core.

### Dragon Dynamic Simulation Model

A simple dynamic model has been developed to simulate the transient behavior of the Dragon Assembly. The goal of this effort is a better understanding of Dragon burst physics. Unlike traditional burst assemblies, such as Lady Godiva[17] where part of the excursion shutdown mechanism is due to increased temperature with the accompanying thermal expansion from fission heating, Dragon's excess reactivity is not quenched by the burst yield, but by the motion of the slug. This model combines the point reactor kinetics model with an energy equation for the fuel and a simple model that describes the motion of the slug. These models are coupled together through a governing assembly reactivity equation and through energy deposition in the fuel. The assembly prompt kinetics is described by the following expression

$$\frac{dP}{dt} = \frac{\beta}{\Lambda}[(R-1)P + \sum_i f_i D_i - R_s] \quad (1)$$

Where

$$f_i = \frac{\beta_i}{\beta} \quad (2)$$

and the source reactivity is given by

$$R_s = -\frac{q}{n_o}\frac{\Lambda}{\beta} \quad (3)$$

The delayed neutron precursor contribution is by the following

$$\frac{dD_i}{dt} = \lambda_i (P - D_i) \quad (4)$$

The reactivity equation for the assembly is given by

$$R = R_s + \alpha_T \Delta T_f + R_{slug} \quad (5)$$

This equation includes a single reactivity feedback mechanism driven by fuel temperature. During drop experiments, the assembly reactivity starts out well below delayed critical, as described by the source reactivity. The excess reactivity generated by the slug passing through the assembly is given by

$$R_{slug} = \frac{2}{\pi} R_{slug,max} \left(\frac{\pi}{2}\frac{h}{a} - \frac{1}{4}\sin\left(\frac{2\pi h}{a}\right)\right) \quad (6)$$

This expression is based on the one-group perturbation model for a simple control rod. The reactivity added to the assembly by the passage of the slug is a function of the total reactivity worth of the slug and the slug's position within the core.

The motion of the slug is described by its free fall from various heights above the core. The instantaneous velocity of the slug is given by

$$v_{slug} = gt \quad (7)$$

This expression assumes that the slug is initially at rest. Therefore, the distance traveled by the slug during its fall is given by

$$L = \frac{1}{2}gt^2 \quad (8)$$

The slug reactivity equation tracks two lengths along the slug's path. The first is when the leading edge of the slug enters the top of the core. This length is given by

$$L_{ct} \leq L \leq (L_{cb} = L_{ct} + a) \quad (9)$$

During this length, the slug's position is described by

$$\frac{dh}{dt} = v_{slug} \quad (10)$$

The second length is when the trailing edge of the slug enters the top of the core. This length is described as follows

$$L_{cb} < L \leq (L_{cb} + a) \quad (11)$$

During this length, the slug position is described by

$$\frac{dh}{dt} = -v_{slug} \quad (12)$$

During the transient, fission energy is deposited into the fuel causing its temperature to increase. The temperature model for the fuel is given by

$$\frac{dT_f}{dt} = \frac{p}{M_f C_p} \quad (13)$$

where the assembly power is given by

$$p = p_o P \quad (14)$$

## *Dragon Criticality Excursion Simulation Results*

The dynamic simulation model described above was implemented in an ordinary differential equation solver and various transient experiments were analyzed. The experiment referred to as Drop No. 73 provides the most useful technical data for simulation. It was estimated that $k_p$ was approximately 1.0035, the average velocity of the slug was 722 cm/s, the initial fission rate of the assembly was $10^5$ fissions/s, and the burst yield was $2 \times 10^{11}$ fissions. It was also noted that the delayed neutron fraction was assumed to be 0.008 and the generation time was estimated to be $1.5 \times 10^{-6}$ s.

The simulation model was run with a source reactivity of -5.90 \$, and a total slug reactivity worth of 7.32 \$. The initial position of the slug was 2.49 m above the top of the core. Both the source reactivity and the total slug reactivity worth fall within the range of the proposed neutronic model parameters that agree with historically measured values. Thus, the dynamic simulation model may be adequately capturing the basic physics of the Dragon Assembly.

At time t=0, the slug begins its fall. The leading edge of the slug reaches the top of the core approximately 0.71 s later. At this point, the slug begins to insert its 7.32 \$ of total reactivity worth. The assembly reaches its maximum excess reactivity of 1.42 \$ at approximately 0.74 s, when the leading edge of the slug reaches the bottom of the core. During this interval, the assembly reaches first prompt critical and the divergent prompt chain reaction begins. The power rises sharply until the leading edge of the slug reaches second prompt critcal, and the power pulse is terminated.

The simulation model predicts a burst yield of $2 \times 10^{11}$ fissions. Figure 11 is a graphical representation of the last 100 milliseconds (of 800 milliseconds) of the simulation results for the Dragon excursion with both first critical and second critical marked on the reactivity curve.

Figure 12 is a graphical representation of the time-dependent integrated power predicted by the simulation model and is plotted similarly to a detector response that records the integrated charge collected as a function of time. This quantity gets more negative as energy is deposited in the detector. This curve has similar characteristics to the integrated pulse curve seen in Figure 4 where the integrated charge collected by the detection system is proportional to the time-dependent integrated energy generated during an excursion. To be able to use the simulation results to re-create a time–dependent detector response curve, the response characteristics of the detector system used back in 1945 is required. From a qualitative perspective, the intergrated power curve has similar temporal and relative intensity characteristics to the detector response curve illustrated in Figure 4.

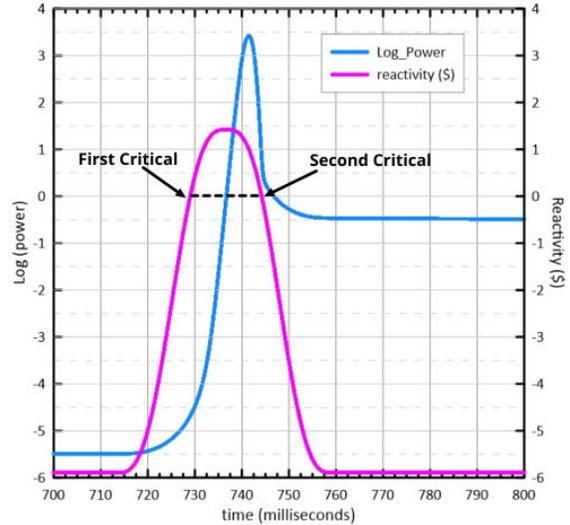

Figure 11. Graphical Representation of last 100 milliseconds for simulation results for the excursion of Dragon Assembly 2.

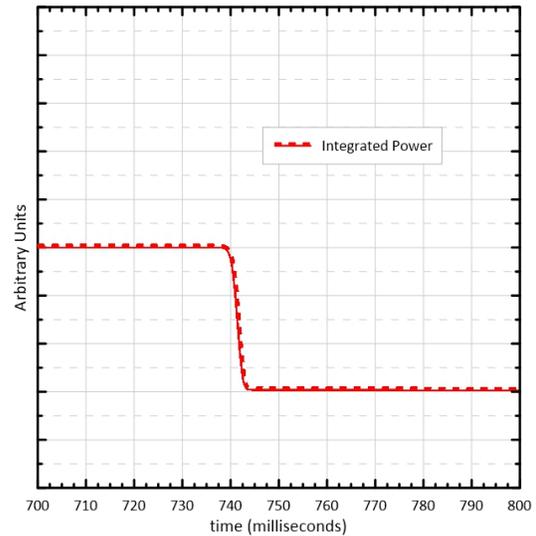

Figure 12. The integrated power is proportional to a detector response that one would observe if using a detection system similar to that used by experimenters during the Dragon excursion measurements.

## *Dragon Index*

$a$ = Height of core (0.165 m)
$\alpha_T$ = Temperature coefficient of reactivity feedback (\$/K)
$\beta$ = Total delayed neutron fraction (0.008)
$\beta_i$ = Delayed neutron fraction of ith group
$C_p$ = Specific heat of fuel (j/g/K)
$D_i$ = Normalized delayed neutron precursor contribution of ith group
$f_i$ = Normalized delayed neutron fraction of ith group
$g$ = Gravitational acceleration (9.8 m/s$^2$)
$h$ = Position of leading edge of slug into core or position of trailing edge in core (m)

$k_p$ = Prompt multiplication factor
$L$ = Distance traveled by slug (m)
$L_{cb}$ = Distance from bottom of core to slug starting point (m)
$L_{ct}$ = Distance from top of core to slug starting position (m)
$\Lambda$ = Mean neutron generation time (1.5x10⁻⁶ s)
$\lambda_i$ = Delayed neutron decay constant for ith group
$M_f$ = Mass of fuel (g)
$n_o$ = Initial assembly neutron population (n)
$P$ = Normalized assembly power
$p$ = Assembly power (W)
$p_o$ = Initial assembly power (W)
$R$ = Assembly reactivity ($)
$R_{slug}$ = Slug reactivity ($)
$R_{slug,max}$ = Total slug reactivity worth ($)
$R_s$ = Source reactivity ($)
$q$ = Neutron source strength (n/s)
$T_f$ = Assembly fuel temperature (K)
$t$ = time (s)
$v_{slug}$ = Slug velocity (m/s)

**Water Boiler Criticality Excursion Simulations**

*Water Boiler Neutronic Model*

Located at Site-Y, the water boiler was described as a homogeneous chain-reacting pile using a solution of uranyl sulphate in water. The fuel solution was contained in a 12" diameter stainless steel vessel. This core was completely surrounded by a reflector of fitted BeO blocks. The mass of Uranium-235 in the spherical core was 565.5 g.

*Water Boiler Dynamic Simulation Model*

As the first liquid-fueled assembly, the water boiler possessed a number of unique characteristics, such as maintaining the core hydrostatically and the ability to alter the fuel composition. A dynamic simulation model has been developed, which combines neutron kinetics with a fuel transport model, to simulate the start-up of the water boiler. The goal of this effort is to illustrate the operating principles of the water boiler and to highlight experimental results. Also, this model may serve as a tool for further study into the behavior of liquid-fueled systems. The model uses the point reactor equations to track the power of the assembly and a simple momentum equation for incompressible fluid flow. The two models are coupled through a reactivity equation that governs the neutronic behavior.

The prompt neutron equation, in normalized form, is given by

$$\frac{dP}{dt} = \frac{\beta}{\Lambda}[(R-1)P + \sum_i f_i D_i - R_s] \quad (15)$$

where the normalized delayed neutron fraction of the ith group is given by

$$f_i = \frac{\beta_i}{\beta} \quad (16)$$

and the source reactivity is given by

$$R_s = -\frac{q}{n_o}\frac{\Lambda}{\beta} \quad (17)$$

The normalized delayed neutron precursor contribution for the ith group is given by

$$\frac{dD_i}{dt} = \lambda_i(P - D_i) \quad (18)$$

The reactivity equation for the assembly during start-up is given by

$$R = R_s + R_{fh} \quad (19)$$

Where $R_{fh}$ represents the reactivity added to the assembly by the addition of fuel solution into the core.

At this point, a number of strategies could be employed to assess the insertion of reactivity as fuel solution gradually fills the water boiler's spherical core. However, in keeping with the spirit of the goal of this model to illustrate the principles of the water boiler's operation, a neutron diffusion theory approach has been chosen. The fundamental one-group criticality condition for a bare homogeneous spherical core may be expressed as

$$k_{eff} = \frac{k_\infty}{1+L_a^2 B_g^2} = \frac{\frac{\nu\sigma_f}{\sigma_a}}{1+L_a^2 B_g^2} \quad (20)$$

Of course, the water boiler was heavily reflected by a close fitting BeO tamper. However, diffusion theory suggests that a reflected spherical core has an "equivalent" larger bare spherical core. The difference in size is described by $\delta$, the "reflector savings", which is simply the difference in radius between the two cores. Given this result, we define the geometric buckling of our water boiler equivalent core as

$$B_g^2 = \left(\frac{\pi}{\frac{h_c}{2}+\delta+l_e}\right)^2 \quad (21)$$

Where $h_c$ is the height of fuel solution in the actual water boiler core. This expression allows us to relate the amount of fuel solution in the core to an estimate of $k_{eff}(h_c)$, as a function of fuel height.

Obviously, a partially filled sphere with a fuel height

$< 2r_c$ is not the same thing as a full sphere with a radius $< r_c$ but, we will leave that issue to the reader's engineering judgement. Given this definition of $k_{eff}$, we may now define the reactivity insertion due to addition of fuel in the core as

$$R_{fh} = \frac{1}{\beta}\left(1 - \frac{1}{k_{eff}(h_c)}\right) \quad (22)$$

Thus, for start-up, the neutronic model only requires the height of the fuel solution in the core. For completeness, we include a reactivity equation for operations performed above delayed-critical. This equation is expressed as

$$R = R_s + R_{fh} + \alpha_T \Delta T_f + \phi V_g + R_{cr} \quad (23)$$

Where $R_{cr}$ describes reactivity insertion due to the control rod. This reactivity equation also accounts for the potential of reactivity feedback due to fuel solution temperature changes and the production of radiolytic gas. This neutronic model now requires additional core parameters. The temperature of the fuel solution may be tracked with a simple adiabatic energy equation given by

$$\frac{dT_f}{dt} = \frac{p}{M_f C_p} \quad (24)$$

where the assembly power is given by

$$p = p_o P \quad (25)$$

A simple radiolytic gas model that tracks the volume of gas within the core is given by

$$\frac{dV_g}{dt} = pG_{rg} - \frac{V_g}{\tau_{rg}} \quad (26)$$

With the neutronic model fully defined, we turn our attention to the hydrodynamic model, which describes the fuel solution handling system. The fuel solution is stored in a shallow inverted right circular cone shaped reservoir. A fill tube running along the axis of the cone connects the fuel solution reservoir to the water boiler core, which is located directly above the fuel solution reservoir. The bottom of the fill tube is located near the apex of the cone and the top of the fill tube enters the core at its south pole. Rubber balloons filled with air provide a cover gas for the fuel solution reservoir. A tube leading from the north pole of the core to a holding tank accommodates the air initially present in the core as it is displaced by fuel solution. By slowly collapsing the rubber balloons, hydrostatic pressure is applied to the fuel solution reservoir. This pressure forces fuel solution from the reservoir, up the fill tube, and into the core. Once the core is filled, the cover gas pressure is held constant. To track the mass flow rate of fuel solution from the reservoir to the core, a simple momentum equation for the fill tube is given by

$$\left(\frac{l_{ft}}{A_{ft}}\right)\frac{dw}{dt} = (P_r - P_c) + \rho_s g l_{ft} - \frac{fl_{ft}}{D_{ft}}\frac{1}{A_{ft}^2}\frac{w^2}{2\rho_s} \quad (27)$$

The pressure of the reservoir near the cone's apex is a function of the reservoir's cover gas pressure and the height of the fuel solution in the reservoir. This pressure is expressed as follows

$$P_r = P_{cgr}(t) + \rho_s g h_r \quad (28)$$

Similarly, the pressure near the south pole of the core is a function of the core's cover gas pressure and the height of fuel solution in the core. This pressure is given by

$$P_c = P_{cgc} + \rho_s g h_c \quad (29)$$

Given the mass flow rate of fuel solution through the fill tube, the change in volume of the reservoir is tracked by

$$\frac{dV_r}{dt} = \frac{-w}{\rho_s} \quad (30)$$

The height of the fuel solution in the reservoir is then expressed as

$$V_r = \frac{\pi}{3}\left(\frac{\sin\theta_R}{\sin\theta_H}\right)^2 h_r^3 \quad (31)$$

Similarly, the change in the volume of fuel solution in the core is given by

$$\frac{dV_c}{dt} = \frac{w}{\rho_s} \quad (32)$$

The relationship between the volume of fuel solution in the core and its height is given by

$$V_c = \frac{\pi}{3}h_c^2(3r_c - h_c) \quad (33)$$

Finally, to account for friction in the fill tube, we may use

$$f = \frac{64}{Re} \quad (34)$$

for laminar flow. For turbulent flow, we may use

$$f = 0.316 Re^{-0.25} \quad (35)$$

for Reynold's number (Re) < 2.0e+04.

Once the core is completely filled with fuel solution, the reservoir cover gas pressure is held constant to maintain the core's fuel solution volume. This hydrostatic pressure condition is given by

$$P_r = P_{cgr}(t_\infty) + \rho_s g h_r \quad (36)$$

Hence,

$$P_r = P_{cgc} + \rho_s g l_{ft} + 2\rho_s g r_c \quad (37)$$

### Water Boiler Manganese Foil Experiment[33]

Among the many noteworthy experiments performed with the water boiler was a Manganese (Mn) foil irradiation experiment. A number of Mn foils were placed at various locations throughout the water boiler assembly and irradiated to map the flux as a function of radial position r. The goal of this experiment was to provide data to compare with the predictions of diffusion and age theory. One interesting result was the determination of the equivalent untamped water boiler core. From diffusion theory, the water boiler experimenters expected the spatial distribution of the flux in an untamped spherical core to be expressed as follows

$$flux \sim \frac{\sin\frac{\pi r}{\mathcal{R}}}{r} \quad (38)$$

Where $\mathcal{R}$ is the radius of the sphere. It was theorized that the spatial distribution of a tamped sphere should be the same as that in a somewhat larger untamped sphere almost to the tamper boundary. From a point near the boundary, a distribution can be extrapolated to give the radius of the equivalent untamped sphere. The Mn foils yielded a ratio of 1.55 between the flux at the center of the core and the flux at the inner edge of the tamper. Given this result, the extrapolation predicted a value of 29.8 cm for $\mathcal{R}$. With this result, the experimenters were then able to calculate the water boiler neutron age as 38.5 cm$^2$, using diffusion and age theory. However, of particular interest to this present study, the determination of the equivalent untamped sphere radius allows us to calculate the reflector savings in our neutronic model. We find that

$$\delta = \mathcal{R} - r_c = 14.6 \; cm \quad (39)$$

The reflector savings[34] is defined as the decrease in the critical core radius when the core is surrounded by a reflector (or tamper).

Years later, Bacher[35] mentioned that Robert Christy was able to make an accurate prediction for the critical mass for a water tamped configuration of a partially enriched uranium solution (Water Boiler). This prediction[36] utilized neutron slowing down diffusion theory and the best available estimates of nuclear data parameters to estimate a critical mass of 600 grams. This prediction was almost exactly the critical mass that was measured experimentally.

### Water Boiler Index

$\alpha_T$ = Temperature coefficient of reactivity feedback ($/K)
$A_{ft}$ = Cross sectional area of fill tube (m$^2$)
$\beta$ = Total delayed neutron fraction
$\beta_i$ = Delayed neutron fraction of ith group
$B_g$ = Geometric buckling (cm$^{-1}$)
$C_p$ = Specific heat of fuel (j/g/K)
$\delta$ = Reflector savings (cm)
$D_i$ = Normalized delayed neutron precursor contribution of ith group
$D_{ft}$ = Diameter of the fill tube (m)
$f$ = Friction factor
$f_i$ = Normalized delayed neutron fraction of ith group
$G_{rg}$ = Radiolytic gas production factor (cm$^3$/W)
$g$ = Gravitational acceleration (9.8 m/s$^2$)
$h_c$ = Height of fuel solution in core (cm)
$h_r$ = Height of fuel solution in the reservoir (m)
$k_{eff}$ = Effective multiplication factor
$k_\infty$ = Infinite multiplication factor
$L_a$ = Diffusion length (cm)
$l_e$ = Extrapolation length (cm)
$l_{ft}$ = Length of fill tube (m)
$\Lambda$ = Mean neutron generation time (s)
$\lambda_i$ = Delayed neutron decay constant for ith group
$M_f$ = Mass of fuel (kg)
$\nu$ = Number of neutrons per fission
$n_o$ = Initial assembly neutron population (n)
$P_c$ = Core pressure at the southpole (Pa)
$P_r$ = Reservoir pressure near the apex (Pa)
$P_{cgc}$ = Pressure of core cover gas (Pa)
$P_{cgr}$ = Pressure of reservoir cover gas (Pa)
$P$ = Normalized assembly power
$p$ = Assembly power (W)
$p_o$ = Initial assembly power (W)
$\phi$ = Void coefficient of reactivity ($/cm$^3$)
$q$ = Neutron source strength (n/s)
$\rho_s$ = Fuel solution density (kg/m$^3$)
$R$ = Assembly reactivity ($)
$R_{cr}$ = Reactivity insertion due to control rod ($)
$R_{fh}$ = Reactivity insertion due to fuel height ($)
$R_s$ = Source reactivity ($)
$r_c$ = Radius of water boiler core (cm)
$\sigma_a$ = Microscopic absorption cross section (cm$^2$)
$\sigma_f$ = Microscopic fission cross section (cm$^2$)
$\tau_{rg}$ = Characteristic residence time of Radilytic gas bubble in core (s)
$\theta_H$ = Angle formed by reservoir slant height and radius

$\theta_R$ = Angle formed by reservoir slant height and height
$T_f$ = Assembly fuel temperature (K)
$t$ = time (s)
$V_c$ = Volume of fuel solution in the core (m$^3$)
$V_g$ = Volume of Radiolytic gas in core (cm$^3$)
$V_r$ = Volume of fuel solution in the reservoir (m$^3$)
$w$ = Mass flow rate of fuel solution (kg/s)

**Dragon Summary**

A number of "ground breaking" results were obtained from the Dragon experiments. From the static experiments, the additive effect hypothesis was established. This operational technique states that the various means that reactivity can be changed have additive effects. This allowed the experimenters to extrapolate from regions of reduced reactivity to above prompt critical. From the drop tests, data was gathered that allowed the experimenters to obtain the neutron generation time. However, the Dragon experiments may be best summarized by Otto Frisch, who wrote "the experiments were done not so much in order to measure any definite quantity, but rather with the idea of demonstrating the existence of divergent chains supported by prompt neutrons only." The Dragon experiments demonstrated a neutron population growing exponentially, doubling in a small fraction of a second. Thus, these experiments proved that a prompt chain reaction possessed an explosive character and military value.

This study presents a dynamic simulation model of the Dragon assembly, in an effort to illustrate the transient behavior occurring during drop tests, in a mathematically rigorous manner. The motion of the slug is coupled to the neutronic behavior of the assembly through an overall reactivity equation. The simulation model's predictions appear to adequately track drop test results. Future work will focus on larger transients that involve fuel heating and reactivity feedback.

**Water Boiler Summary**

In addition to the Mn foil irradiation experiments described above, the water boiler provided a wealth of data and experience on reactor operations. As an example, the Inhour relationship obtained on the water boiler was then used for the Dragon[37,38] assembly. Many of the routine reactor operations performed today, such as approaches to critical, control rod worth calibrations, and power calibrations, were used extensively to test alternate reflector materials and fuel compositions.

Perhaps the most remarkable aspect of the water boiler is its legacy at Los Alamos. The water boiler was the progenitor of LOPO, HYPO, SUPO, Kinglet[39], and SHEBA[40,41]. The simplicity and versatility of these machines were demonstrated by a wide range of applications in the areas of criticality safety, intense radiation environments, and medical isotope production. At present, Los Alamos continues to develop novel liquid-fueled reactor concepts, such as the Pumped Fuel Aqueous Homogeneous Reactor (PFAHR)[42]. See Figure 6. The PFAHR circulates a fissile solution around a loop to achieve large power outputs. The dynamic simulation model presented above served as a design tool in the development of the PFAHR. Future work will focus on novel applications of the PFAHR in the areas of medical isotope production and intense radiation environment testing.

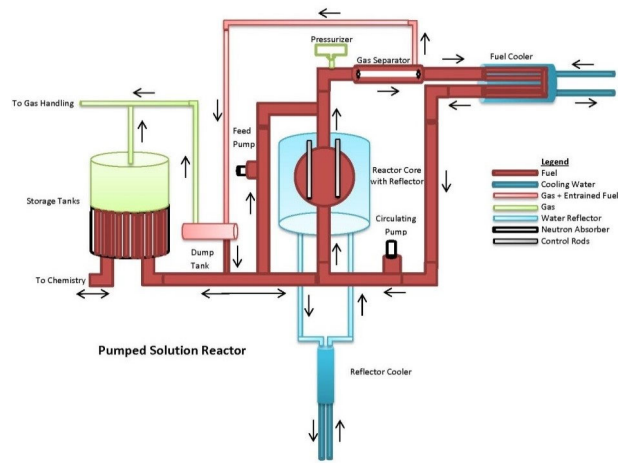

Figure 13. Pumped Fuel Aqueous Homogeneous Reactor (PFAHR) Conceptual Design Schematic

**Acknowledgements**

The authors would like to thank ALD-GS for providing some financial resources to complete this work. Also, the authors would like to provide a special thanks to Alan Carr and Daniel Alcazar in the Weapons Research Services/Secure Information Sevices group (WRS-SIS) for helping us search and navigate the Laboratory's archival holdings of documents, logbooks, drawings, and photographs related to the Manhattan project records that were pertinent to this work.

This work was supported by the US Department of Energy through the Los Alamos National Laboratory. Los Alamos National Laboratory is operated by Triad National Security, LLC, for the National Nuclear Security Administration of the US Department of Energy under Contract No. 89233218CNA000001.